# Visible Light Spectroscopy of Liquid Solutes from Femto- to Attoliter Volumes inside a Single Nanofluidic Channel

*Björn Altenburger[1], Joachim Fritzsche[1] and Christoph Langhammer[1*]*

[1]Department of Physics, Chalmers University of Technology; SE-412 96 Gothenburg, Sweden

*Corresponding author: clangham@chalmers.se

**Abstract**

UV-Vis spectroscopy is a workhorse in analytical chemistry that finds application in life science, organic synthesis and energy technologies like photocatalysis. In its traditional implementation with cuvettes, it requires sample volumes in the milliliter range. Here, we show how Nanofluidic Scattering Spectroscopy, NSS, which measures visible light scattered from a single nanochannel in a spectrally resolved way, can reduce this sample volume to the attoliter range for solute concentrations in the mM regime, which corresponds to as few as $10^5$ probed molecules. The connection of the nanochannel to a microfluidic in-and outlet system enables such measurements in continuous flow conditions, and the integrated online optical reference system ensures their long-term stability. On the examples of the non-absorbing solutes NaCl and $H_2O_2$, and the dyes Brilliant Blue, Allura Red and Fluorescein, we demonstrate that spectral fingerprints can be obtained with good accuracy and that solute concentrations inside the nanochannel can be determined based on NSS-spectra. Furthermore, by applying a reverse Kramers-Kronig transformation to NSS-spectra, we show that the molar extinction coefficient of the dye solutes can be extracted with excellent agreement with the literature values. These results thus advertise NSS as a versatile tool for the spectroscopic analysis of solutes in situations where nanoscopic sample volumes, as well as continuous flow measurements, are critical, e.g., in single particle catalysis or nanoscale flow cytometry.

## Introduction

Spectroscopy based on electromagnetic radiation is one of the most fundamental experimental principles in modern science and has enabled an uncountable number of advances in research since its formal description[1] by Isaac Newton in 1704. For example, it has offered invaluable insights into the nature of light and matter due to a plethora of interactions between different mechanical or electromagnetic frequencies, and the very basic structures of, e.g., solid crystals, molecules, atoms, and electrons. Investigating the frequencies of these interactions individually, i.e., in a spectrally resolved manner, yields information about structure, composition, electronic transitions and chemical reactions that occur inside or on the surfaces of the sample the radiation interacted with or originated from. Because of this wide applicability and the wide range of information that can be obtained, spectroscopy based on electromagnetic radiation has become a workhorse in the experimental method toolbox of the natural sciences and has since its introduction diversified into a plethora of associated methods and techniques.

To structure these spectroscopic methods, we can organize them according to the size/volume of the investigated sample. To this end, the standard versions of spectroscopic tools typically enable investigations of macroscopic samples only, that is, samples that are contained in relatively large containers, such as cuvettes or gas chambers, or samples like crystals and powders whose volumes are large compared to the actual molecules or particles of interest. These techniques can then be ordered by increasing frequency (or decreasing wavelength) of the electromagnetic radiation exploited to probe the sample, that is, for instance, microwaves[2], infrared radiation[3–6] (IR), ultraviolet-visible (UV-Vis) light, X-rays[7,8] and electrons[9,10].

While all very powerful in their specific regime, methods employing UV-Vis light have earned a special place in this list since, e.g., many natural processes require the interaction of sunlight with (living) matter, which means that many biologically important molecules absorb UV-Vis light[11–14]. Furthermore, it is also the same wavelength range that is used for modern optical communication[15,16] and as source for sustainable energy[17–21]. The main methods used in this spectral domain are absorption[22] and emission spectroscopy[21,23–25], surface plasmon resonance (SPR)[26] and localized surface plasmon resonance (LSPR)[27] spectroscopy, as well as Raman spectroscopy[28]. They, e.g., can reveal the electronic structure of molecules and transitions between different electronic states or electronic excitations and transitions in solids, such as metals and semiconductors, as well as small refractive index (RI) changes induced, e.g. by the specific binding of molecules onto plasmonic surfaces.

Considering the typical dimensions of the entities that interact with the radiation at hand, efforts to reduce the required sample size or volume to the micro- or even nanoscale have been a strong driving force in the development of new spectrometric methods. As key reasons, we identify that reducing the required amount of sample material is desirable both from a cost perspective when expensive/scarce raw materials are used, and since the synthesis of large volumes of new materials or molecules often is very difficult and time consuming, and hence hampers the throughput in, e.g., screening processes. Furthermore, spectroscopic methods that provide micro- or nanoscopic spatial resolution (enabled by tiny sample volumes being enough to generate a measurable signal) make it possible to investigate nanomaterials, particles, surfaces and molecules in greater detail, in a manner that can differentiate between locations and species, and ideally also beyond ensemble averaging. These prospects have driven the development of nanoscale spectroscopy methods[29], such as IR atomic force (AFM) spectroscopy[30,31], IR scanning tunneling microscopy (STM) spectroscopy[32,33], tip-enhanced Raman-spectroscopy[34], scanning near-field optical microscopy[35] (SNOM), electron energy loss spectroscopy (EELS), cathodoluminescence methods[35], coherent phonon spectroscopy[36], single particle plasmonic spectral sensing[37], as well as many approaches involving fluorescence[38,39]. It is evident from this list that many of these techniques either employ tip- or plasmonic nanoparticle/surface-based signal enhancement concepts often in conjunction with scanning microscopy and/or utilize the IR or ultrashort wavelength regime (X-ray, electrons) of the electromagnetic spectrum. Especially noteworthy for our discussion here are concepts that integrate surface enhanced Raman spectroscopy (SERS) with nanofluidics, as done with colloidal nanoparticles[40] or nanoslits[41], enabling the detection of protein folding states and single nucleobases, respectively. Furthermore, the use of nanoscale hotspots in conjunction with surface-enhanced infrared absorption (SEIRAS) made it possible to quantify molecules in a nanofluidic system[42]. While these results clearly demonstrate the possibilities of fluorescence and surface-enhanced Raman and IR approaches, we identify a distinct lack of nanoscale spectroscopy methods that are label-free, i.e., do not require fluorescence, and that do not require signal enhancement by plasmonic tips, surfaces or nanoparticles. An interesting approach that relies only on the intrinsic properties of molecules to decrease the detection limit while omitting surface enhancements is given by photothermal optical diffraction (POD)[43] and photothermal optical phase shift (POPS)[44], both relying on the temporary variance in refractive index caused by the thermal de-excitation of molecules. However, these techniques do not provide spectral analysis in the Vis spectral range, making it clear that a label and plasmonic enhancement-free nanofluidic characterization method for this wavelength range is missing. This means that there is to our knowledge currently no method for nanofluidic applications that reproduces the performance of established macroscopic UV-Vis spectroscopy.

To fill this gap, we present Nanofluidic Scattering Spectroscopy (NSS) that enables label-free spectroscopy in the visible light (Vis) regime inside individual nanofluidic channels. These channels offer tiny sample volumes, in our study here ranging from 2 femtoliter (fl) to 60 attoliter (al), depending on chosen nanochannel dimensions and settings of the CCD camera. This concept is a further development of Nanofluidic Scattering Microscopy (NSM) we recently have introduced[45,46]. As the key features beyond the state of the art, NSS is not only able to deliver the full wavelength-dependent scattering and absorption spectrum, and thus the wavelength-dependent molar extinction coefficient of a solution contained inside a single nanofluidic channel, but also to reveal its wavelength-dependent RI and the absolute concentration of solute, as we illustrate on five examples of transparent, colored and fluorescent molecular solutions in the millimolar to molar concentration regime. Furthermore, in contrast to similar techniques on open surfaces[47], since a nanofluidic channel connected to a microfluidic system is used as sample vessel, continuous convective flow is enabled and thus facilitates both high experimental throughput and the possibility to monitor changes in the probed solution continuously.

**Results and discussion**

Scattering of electromagnetic radiation from small objects or molecules is a phenomenon often encountered in nature. It is the reason why the sky is blue but also why tiny scratches in a window are visible. Mechanistically, those scratches scatter light because they are objects with a different RI than the surrounding glass (n = 1.459[48]) since they are filled with air (n = 1), and because they are smaller (Rayleigh scattering) or comparable (Mie-scattering) to the wavelength of the irradiated light. In analogy to scratches in glass, in this work, we make use of the light scattered from nanofluidic channels etched into a silicon dioxide ($SiO_2$) surface and hermetically sealed with a glass lid (**Figure 1a**). The choice of $SiO_2$ as substrate is well established within nanofluidics[45,46,49], as it offers great chemical resistance, and various fluidic designs can be realized using fabrication methods established for silicon-based microelectronics[50]. Specifically, the nanochannels we use here constitute an elongated rectangular cavity with 200 nm width and depth, and 62 micrometers length, nanofabricated into the 250 nm thick thermal oxide layer of a silicon wafer (**Figure 1b**). The inset in **Figure 1b** depicts a SEM-image of the nanochannel cross section and reveals the nearly rectangular shape with small features that are the consequence of the used etching process. These features do not impact the scattering properties of the channel, as we have corroborated in our earlier work by comparing an analytical model of cylindrical nanochannels with exact electrodynamic simulations of nanochannels with square cross sections[51]. We also note that principally, the exact cross-sectional dimensions, as well as the nanochannel length, can be tailored within a wide range and according to the needs of a specific measurement since they are crafted using highly flexible micro- and nanolithography techniques described in detail in the Methods section. By connecting such nanofluidic channels to a microfluidic in- and outlet system described in more detail below, it becomes possible to flush liquid solutions[52,53] but also gases[54] through the channel and thereby change the RI difference between the solution in the channel and the surrounding medium, i.e., $SiO_2$ in our current design. This change, in turn, alters the spectral distribution and intensity of visible light scattered from the nanochannel. Detecting these spectral changes constitutes the operation principle of NSS.

*Theoretical foundation of NSS*

To describe the NSS principle on a more analytical basis, we approximate the rectangular nanochannel from the experiment as a cylinder of infinite length (**Figure 1c**). The validity of this approximation has been corroborated previously using finite-difference time-domain (FDTD) simulations[55]. Furthermore, we adopt the description of the propagation of unpolarized incident electromagnetic waves and how they interact with a nanochannel from Bohren and Huffmann[56] to arrive at an expression that describes the scattering cross section of a nanochannel, $\sigma_{channel}$, as a function of the geometrical channel cross section, $A_\emptyset$, its (illuminated) length, $L$, and the wavenumber of the incident light, $k = 2\pi/\lambda$, as[46]

$$\sigma_{channel} = \frac{A_\emptyset^2 k^3 L}{4}(m^2 - 1)^2 \left(\frac{1}{2} + \frac{1}{(m^2+1)^2}\right). \qquad \text{Equation 1}$$

This expression also contains the parameter $m = n_l/n_{SiO2}$, which is the ratio of the RIs of the liquid in the channel, $n_l$, and of the surrounding medium of the channel, which is SiO$_2$ in the present case, $n_{SiO2}$. It thus becomes evident that a change in $n_l$ indeed induces a change in the nanochannels´ scattering cross-section, provided all other variables remain constant.

As the first step to establish the fundamental understanding of the NSS methodology, it is relevant to analyze the wavelength-dependence of the RI of SiO$_2$[48] (channel matrix) and water[57] (solvent used), as well as of their ratio, $m$ (**Figure 1d**). Evidently, both RIs only weakly depend on the wavelength and in a very similar way, such that $m$ remains fairly constant across the visible wavelength range. Furthermore, calculating the scattering cross sections across the visible spectral range for a water-filled nanochannel for parallel ($\sigma_p$), orthogonal ($\sigma_o$) and unpolarized (average, $\sigma_{channel}$) incident light reveals that its scattering cross section increases significantly for shorter wavelengths and most strongly for parallel polarization (see **Figure 1d**).

To illustrate how the optical contrast is generated in NSS, it is interesting to plot how the scattering cross section of the channel depends on $n_l$ over a large $n_l$-range for a given wavelength (here 600 nm, since it is the center wavelength of the spectrometer we use in our experiments, see **Figure 5a**), and again for parallel, orthogonal and average polarization of the incident light (**Figure 1e**). A complete analysis for all wavelengths is given in the Supplementary Information (**Figure S2**). This reveals that scattering vanishes when $n_l = n_{SiO2}$ and that it increases in an almost quadratic fashion when $n_l \leq n_{SiO2}$, such that, e.g., an air filled channel ($n = 1$) scatters much more light than a water filled channel ($n = 1.333$). We also see that even very small RI changes induce a sizable change in the scattering cross section, as we have exploited in our previous work to quantify concentration changes induced by a catalytic reaction on single nanoparticles[46].

For the reverse scenario, $n_l \geq n_{SiO2}$, it is evident that the scattering cross section increases even more quickly per change of RI (**Figure 1e**). However, liquids with an RI larger than $n_{SiO2}$ are scarce and not commonly used as solvents. Nonetheless, this fact is interesting for future development since a nanofluidic system embedded into a matrix with a lower RI than the liquid inside it would boost the scattering intensity of the system and thus its ability to discern small RI contrasts.

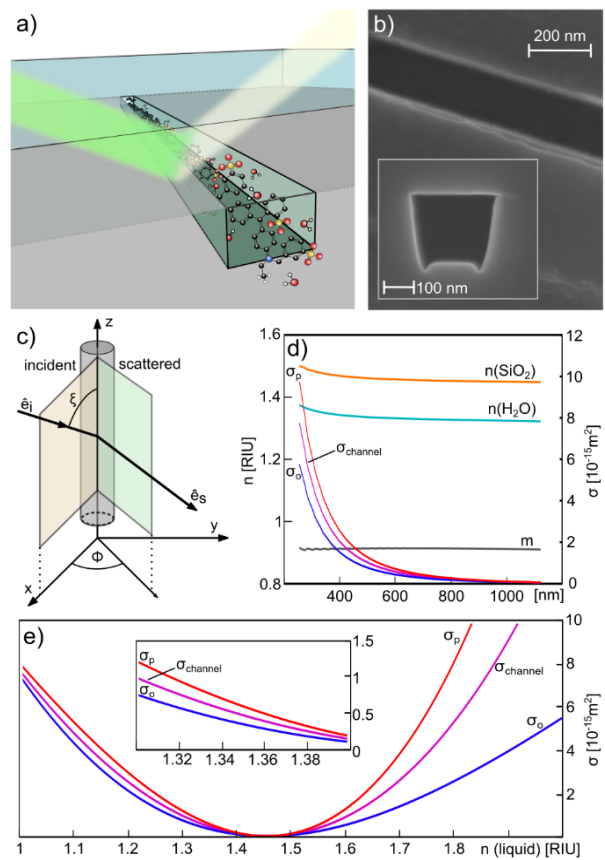

*Figure 1. Principle of Nanofluidic Scattering Spectroscopy. a) Artistic rendering of a nanochannel filled with dye molecules. An incident light beam is partially absorbed by the dye molecules, partially scattered on the channel, thereby changes its spectral composition (see **Figure S1** for a complete sketch of the experimental setup). b) SEM image of a 200 nm x 200 nm nanochannel. The inset shows a cross*

*section of a nanochannel, being nearly rectangular. The small features are a consequence of the etching process. c) Sketch of the analytical theoretical model used, with the nanochannel approximated as cylinder (grey) and the planes of the incident (orange) and scattered light (green). The Poynting vector of the light is shown in black. d) Incident light wavelength, λ, dependence of the scattering cross section, σ, of a water-filled channel shown for a polarization parallel to the channel ($σ_p$), orthogonal to it ($σ_o$) and for unpolarized light ($σ_{channel}$). Data plotted together with the wavelength dependence of the RIs, n, of water[57] and silicon oxide[48] and their ratio, m. e) Calculated scattering cross section of a 200 nm by 200 nm channel embedded in silicon oxide (n = 1.459) as a function of the RI of a medium inside the channel, again shown for three different polarizations of the incident light at 600 nm (see **Figure S2** for all visible wavelengths). The inset shows a zoom in of the RI region relevant for aqueous solutions.*

Having established the theoretical foundation, it is now interesting to consider some first practical aspects. Specifically, we note that what is recorded in an NSS experiment is the spectrally resolved intensity of light scattered from a nanofluidic channel, that is, a scattering spectrum. In contrast, however, in traditional UV-Vis spectroscopy – which we here call absorption spectrophotometry (ASP) - an absorption (or absorbance) spectrum is measured and enables the quantification of, e.g., molar extinction coefficients, which constitute a fundamental material constant. It is therefore of interest to introduce a general formalism, here first detached from nanochannels, that enables the mathematical transformation of an absorption spectrum into a scattering spectrum.

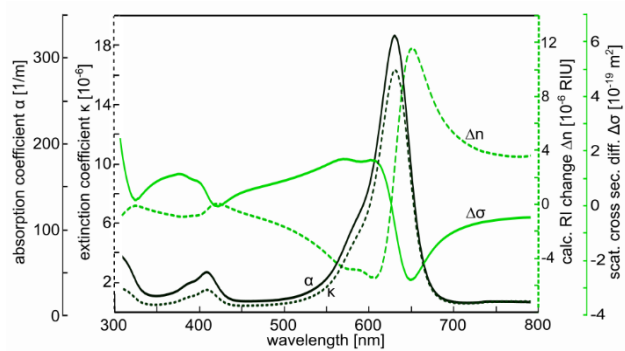

*Figure 2. Analytical transformation from absorption to scattering spectrum for Brilliant Blue. We start with the spectrum of the absorption coefficient, α, for Brilliant Blue (black solid line) as obtained by ASP. In the first step, this spectrum is converted to the spectrum of the extinction coefficient, κ, (black dashed line). Subsequently, the corresponding change in RI spectrum, Δn (green dashed line) is calculated using the Kramers-Kronig relation. Finally, the change in RI spectrum is converted to a scattering cross section spectrum difference, Δσ, of the nanochannel (green solid line). The corresponding transformations for Allura Red and Fluorescein are depicted in **Figure S3**.*

To do this, we first remind ourselves that the RI of a solid or liquid is a complex function of wavelength as

$$\boldsymbol{n}(\lambda) = n(\lambda) + i\kappa(\lambda). \qquad \textit{Equation 2}$$

Secondly, we consider the absorption coefficient spectrum, $\alpha(\lambda)$, using the Brilliant Blue dye (**Figure 2**, black solid line) measured by standard UV-Vis spectroscopy as example. It can be converted into an extinction coefficient spectrum, $\kappa(\lambda)$, (**Figure 2**, black dashed line) using the expression

$$\kappa(\lambda) = \frac{\alpha(\lambda)\,\lambda}{4\pi}. \qquad \textit{Equation 3}$$

Subsequently, since the extinction coefficient corresponds to the imaginary part of the RI of the Dye (**Equation 2**), we can calculate the spectral variation in the real part of the RI induced by absorption, $\Delta n(\lambda)$, (**Figure 2**, green dashed line) using the Kramers-Kronig relation[58,59,60] as

$$\Delta n(\lambda) = n(\lambda) - 1 = \frac{2}{\pi} \mathcal{P} \int_0^{-\infty} \frac{\kappa(\lambda')}{\lambda'\left(1-\left(\frac{\lambda'}{\lambda}\right)^2\right)} d\lambda'. \qquad \textit{Equation 4}$$

Finally, substituting the obtained expression for the wavelength-dependent RI, $n(\lambda)$, of the dye into **Equation 1**, we can connect the above analysis to the nanochannel framework. Accordingly, we can now calculate the expected change of the scattering cross section of a nanochannel, $\Delta\sigma$, (**Figure 2**, green solid line) induced by a change of the liquid, and thus the RI inside the channel, which may be induced by a change in the concentration of a solute or the complete exchange of said liquid.

*Nanofluidic design*

For the NSS experiments, we designed a fluidic chip schematically depicted in **Figure 3a** and further illustrated by corresponding dark-field scattering microcopy images taken at different magnifications (**Figure 3b-d**). As a key step beyond the state of the art[45,46], we have implemented two independent fluidic systems on the chip. The first one serves as the sample system (**Figure 3a** - blue) through which different sample solutions are introduced. The second one constitutes a reference system (**Figure 3a** - orange) filled with water at all times that we will use to compensate for, e.g., fluctuations in irradiated light intensity during a measurement. The sample solution exchange is enabled through a system of microchannels (50 μm wide and 1.2 μm deep), which at the inlet are connected to a macroscopic reservoir via an O-ring seal (**Figure 3e-f**), and which on the other end are connected to an array of parallel nanochannels with cross-sectional dimensions of 200 nm x 200 nm via a smaller microchannel (2 µm wide and 1.2 µm deep).

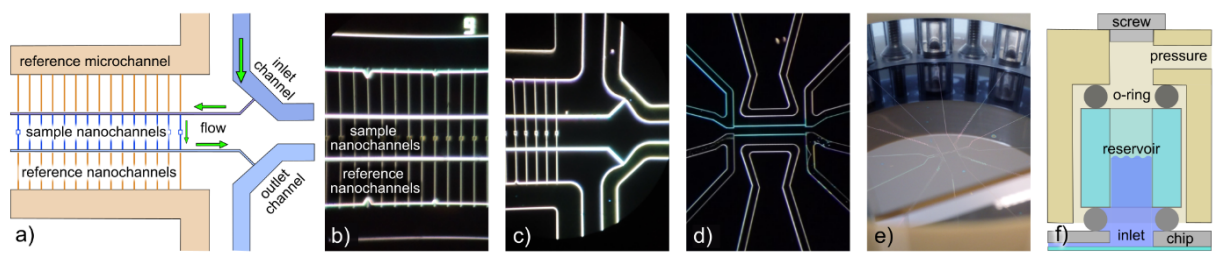

*Figure 3. Nanofluidics layout.* a) Schematic layout of the two micro- and nanofluidic systems implemented on the chip. The reference nanochannel system used for online optical referencing (orange) has nanochannels that are arranged colinearly with the sample nanochannels in the center (blue). Importantly, however, the two systems are not physically connected to ensure that the reference channel system always is filled with the desired reference liquid, i.e., the solvent – water in the present case, while the liquid is exchanged in the sample fluidic system during a measurement. Liquid access and exchange in the nanochannels are enabled via access microchannels (light blue for the sample system and orange for the reference system) by applying higher pressure to the inlet channel side. b) Dark-field scattering microscopy image of the sample and reference nanochannels. The distance between the nanochannels is 20 µm and the sample channels are 62 µm long with a geometric cross section of 200 nm x 200 nm (*cf. Figure 1b*). The bright points in the center of the sample channels are constrictions that enable the trapping of colloidal nanocrystals, as we have demonstrated in earlier work[45,46]. Their function is not used in this work. c) Dark-field scattering image of the interface between micro- and nanofluidic systems. d) Dark-field scattering image of the center of the fluidic chip at a lower magnification. Note the blue hue of the sample fluidics system as it has been filled with Brilliant Blue solution. e) Photograph of the fluidic chip installed in the chip holder. The microfluidic systems are visible and connected to liquid reservoirs in the outer perimeter of the chip holder. f) Schematic cross section of the inlet reservoirs showing how the chip holder connects to the chip.

The same arrangement is mirrored on the outlet side of the chip. The reference nanochannels, which have the same cross-sectional dimensions as the sample channels to be optically identical, are arranged in between and parallel to each of the sample channels, such that one sample and one reference channel fit within the opening of the slit of the spectrometer during an experiment (**Figure 4g**).

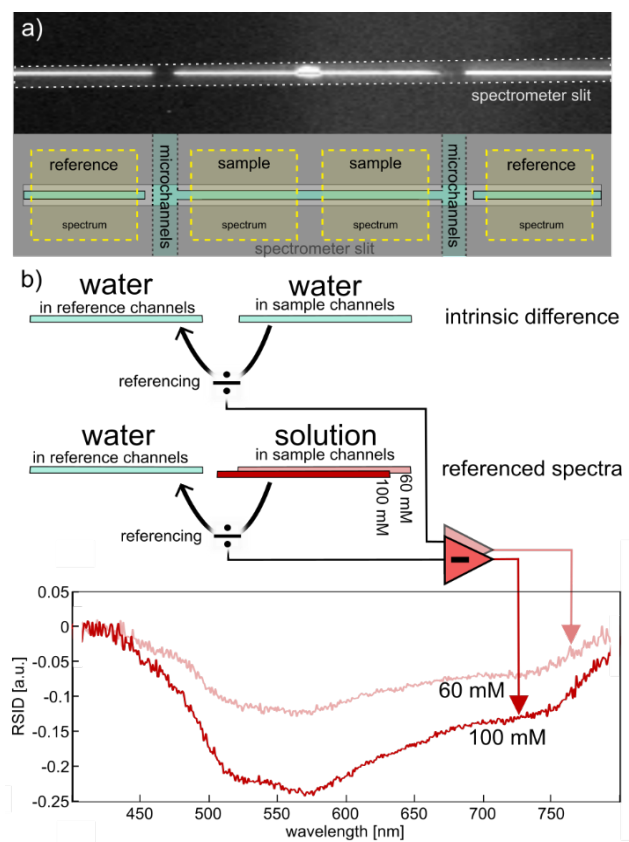

***Figure 4. On-chip online optical referencing.*** *a) To enable the simultaneous online acquisition of sample and reference scattering spectra, we align a set of sample and reference nanochannel with the slit of the spectrometer. The nanochannels are sized such that spectra from to two separate reference channels and from two separate areas of the sample channels can be recorded simultaneously (areas marked with yellow). The bright area in the center of the sample channel is again a constriction for colloidal particle trapping not used in this study. The microchannels are not visible due to the illuminating light being incident parallel to the microchannel walls. b) Graphic depiction of the optical referencing scheme that delivers the relative scattering intensity difference (RSID) spectra: i) both sample and reference nanochannels are filled with solvent water and the spectrum obtained from the sample channel is divided by the spectrum from the reference channel to deliver the intrinsic scattering intensity difference spectrum between sample and reference channel. ii) the sample channel is filled with the solution of choice and the corresponding scattering spectrum from the sample channel is divided by the (constant) reference spectrum from the still water filled reference channel. iii) The intrinsic scattering intensity difference spectrum obtained in step i) is subtracted from the normalized sample spectrum to obtain the RSID spectrum. c) RISD spectra obtained by this procedure for 100 mM and 60 mM Allura Red solutions.*

*Chip-integrated continuous optical referencing*

The concept of continuous online optical referencing is a very efficient way to reduce noise and drift induced by fluctuating light intensities, change of focus of the microscope or thermal (expansion) induced effects, and it is widely applied in dual-beam spectrophotometers. To implement such online referencing in an NSS experiment, we rely on the set of reference channels introduced above and the signal treatment sequence depicted in **Figure 4a-b** that comprises the following steps.

(i) Both the reference and sample channel systems are filled with water (or any other solvent used for a specific experiment) by applying a pressure of 2 bar to the corresponding reservoirs in the chip holder. A nanochannel pair is placed in the slit of the spectrometer and four scattering spectra are recorded, two from the reference channel at an upstream and downstream position, and two from the sample channel at the same up- and downstream position. It is here of critical importance that the illumination and observed channel length are as identical as possible for all positions, because the signal ratio of sample and reference spectra during this step needs to be close to 1 for the evaluation scheme to be valid (see also **Figure S4**). One of the two sample-reference spectra pairs is used as a backup and control, as it may happen that sample solution is leaking into the reference channels or that the channels are compromised in another way. To record these spectra, we bin the signal from 50 pixels along the respective nanochannel and position, which corresponds to a ca. 30 µm long fraction of the respective nanochannel, to reduce noise. Subsequently, we divide the obtained water-filled sample channel spectrum by the water-filled reference channel spectrum to obtain a what we call "*intrinsic difference spectrum*" between sample and reference channel. We will use this intrinsic difference spectrum (examples shown in **Figure S4a**) in the last analysis step (iii) to account for the intrinsic differences in scattering profile that (may) exist between a reference and sample channel when they are filled with the same liquid, e.g., due to slightly different dimensions or surface roughness.

(ii) The water in the sample channel system is exchanged by an aqueous solution of the compound of interest (here the dye Allura Red, as a first example), by exchanging the liquid in the corresponding reservoir and again applying 2 bar of pressure to establish a flow through the nanochannel. The measured scattering spectrum from the sample channel is then divided by the simultaneously obtained spectrum of the water filled reference channel.

(iii) As a final step, the intrinsic difference spectrum measured in step (i) is subtracted from the referenced sample spectrum obtained in step (ii), resulting in what we call a *"relative scattering intensity difference"* (RSID) spectrum.

*NSS measurements of non-absorbing solutes*

To illustrate the principle of NSS and its application as a method for the detection of solute concentrations inside a nanofluidic channel, we consider two solutes that are transparent, i.e., do not absorb light in the UV-Vis regime: NaCl and $H_2O_2$ dissolved in water (**Figure 5**). To establish these experiments, it is illustrative to first discuss the irradiated spectrum produced by the used LED white light source and how it is affected by the optical elements in the experimental setup comprised of microscope, spectrometer and CCD camera, to understand the origin of specific features in the spectra obtained in our experiments. The light emitted from the LED has a relatively broad emission band spanning from 420 nm to 760 nm, with the highest intensity around 540 nm. In addition, there is a strong peak at 450 nm (**Figure 5a**). However, when this light is scattered from a water-filled nanochannel, the maximum intensity of the scattered spectrum is significantly shifted when measured through our microscope system (**Figure 5a** and **Figure S5**). This is the consequence of (i) the spectral sensitivity of the CCD camera, (ii) the wavelength-dependent efficiency of the used grating in the spectrometer and (iii) the transmittance characteristics of other optical elements on the microscope, such as the objective (**Figure S1**). Together, they skew the emission spectrum of the lamp to longer wavelengths and thus, e.g., strongly decrease the 450 nm peak in the irradiated spectrum. However, when exchanging the water in the sample channels to a 5M NaCl solution, the overall shape of the scattered light spectrum obtained from a sample volume of about 1.2 fl (corresponds to the 30 μm long nanochannel section used) remains globally very similar, since there are no absorption bands for the NaCl solute (**Figure 5a**). Nonetheless, we notice sizable differences in scattered light intensity that are most pronounced around 600 nm, which we can ascribe to the presence of the solute and the corresponding change in RI of the solution.

As the next step, we extracted RSID spectra for seven different NaCl concentrations ranging from 0.25 M to 5 M in water, according to the procedure introduced in the previous section (**Figure 5b**). Due to the lack of absorption bands of this solute, the obtained RSID spectra are broad and featureless, with a strictly negative amplitude due to an increasing RI compared to pure water upon increasing solute concentration. This amplitude is proportional to the solute concentration, as further quantified below. Similar results are obtained when using $H_2O_2$ as the solute, however, with sizably different RISD magnitudes (**Figure 5c**). These measurements demonstrate that concentration-dependent UV-Vis scattering spectra of a non-absorbing solute can be obtained from 1.2 femtoliter sample. If we consider here the highest concentration of each solute, the number of molecules sampled in **Figure 5b-c** is $3.6 \cdot 10^9$ for NaCl and $7.2 \cdot 10^9$ for $H_2O_2$, and for the lowest concentration it is $1.8 \cdot 10^8$ and $3.6 \cdot 10^8$, respectively.

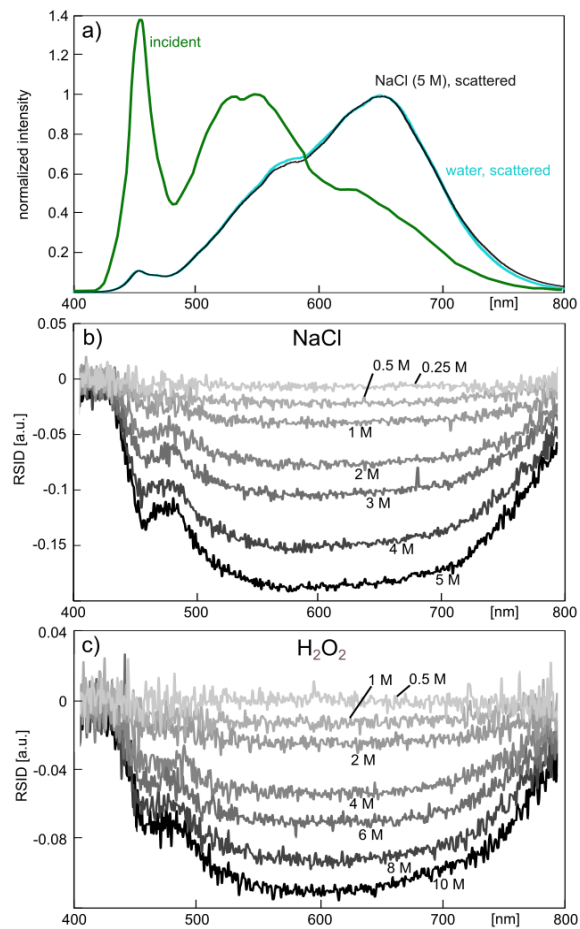

***Figure 5. NSS of NaCl and $H_2O_2$ solutions.*** *a) Incident emission intensity spectrum of the LED light source prior to entering the microscope (green), this light scattered from a water-filled nanochannel measured through the NSS microscope setup (cyan), and this light scattered from a nanochannel filled with a 5 M NaCl solution measured through the NSS microscope setup (black). All spectra have been*

*normalized to their maximum value in the wavelength range between 500 nm and 700 nm. RSID Spectra for b) NaCl and c) $H_2O_2$ at different concentrations in water, as indicated by the labels.*

*NSS measurements of light-absorbing and fluorescent dyes*

Since traditional UV-Vis spectroscopy is widely applied to determine solute concentrations, as well as characteristic fingerprints of molecules that exhibit absorption bands in the UV-Vis spectral range, we in the next step apply NSS to a selection of organic and fluorescent dyes, i.e., Brilliant Blue, Allura Red and Fluoresceine, in the concentration regime from 5 mM to 100 mM.

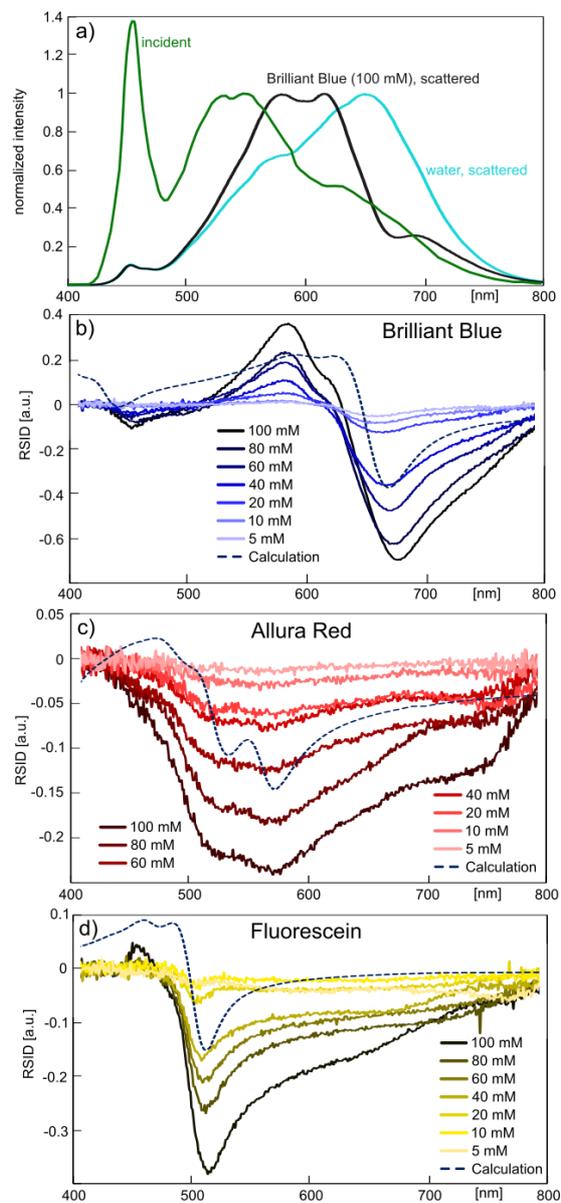

*Figure 6. NSS of Brilliant Blue, Allura Red and Fluorescein solutions.* *a) Incident emission intensity spectrum of the LED light source measured directly, i.e. not through the microscope setup (green), this light scattered from a water-filled nanochannel measured through the NSS microscope setup (cyan), and this light scattered from a nanochannel filled with a 100 mM Brilliant Blue solution measured through the NSS microscope setup (black). All spectra have been normalized to their maximum value in the wavelength range between 500 nm and 700 nm. RSID Spectra for b) Brilliant Blue, c) Allura Red and d) Fluorescein at different concentrations in water, as indicated by the legend. Each panel also includes a scaled back- calculated scattering spectrum obtained from ASP measurements of the respective dye solution.*

Starting again at the level of the raw measured scattering spectra and taking a 100 mM Brilliant Blue solution as the example, we notice that the distinct absorption bands of the dye are clearly reflected when comparing the spectra obtained from a water-filled and a dye-filled nanochannel (**Figure 6a**). Subsequently deriving the RSID spectra for the 1.2 fl sample volume (30 µm long nanochannel section) for the entire concentration range reveals a systematic dependence of RSID intensity on concentration, as well as a distinct peak at 580 nm with a weaker "shoulder" at 630 nm, and a distinct negative peak at 680 nm (**Figure 6b**). These features correspond to rising and falling flanks in the absorption spectra of the dye (cf. **Figure 2, Figure S3)** and can be reasonably reproduced when calculating theoretically the expected scattering spectrum of Brilliant Blue inside a nanochannel using the formalism derived above, i.e., **Equation 4** (dashed line in **Figure 6b,** see also **Figure S3**). The observed discrepancies between calculated and measured RSID spectra are most likely the consequence of the significant spectral modulation of the irradiated light intensity in our setup (cf. **Figure 5a** and corresponding discussion), which enhances or suppresses certain features relative to each other, depending on their spectral position. As a last point, we notice that for Brilliant Blue both positive and negative RSID amplitude changes can be observed, because of positive or negative RI-differences induced by the different absorption bands compared to pure water.

Executing the same analysis for the dye Allura Red (**Figure 6c**) and the fluorescent dye Fluorescein (**Figure 6d**) in the same concentration range and the 1.2 fl sample volume generates overall very similar results with reasonable agreement between calculated (see **Figure S3**) and measured RSID spectra, and with distinct concentration dependence of the RSID signal amplitudes. We note that we don't resolve the slight positive RSID peak at for Allura Red predicted by the calculated spectrum. This is the consequence of the dye's small extinction coefficient and the low transmittance of our microscope setup in the wavelength range where the peak is expected to occur, as evident from **Figure 6a**. We also note that for Fluorescein, we do not resolve a scattering band that would correspond to the 540 nm fluorescence emission line. The likely reason is that the number of emitted fluorescence photons is much lower than the number of scattered photons from the nanochannel since the number of fluorescein molecules in the channel at the given concentrations is very low, i.e., on the order of $10^7$ molecules.

To put this number into perspective and compare NSS in this respect to typical solute concentrations and sample volumes used in ASP, we recall that in standard ASP cuvettes, with an optical path of 1 cm through the sample solution, the commonly used solute concentrations range from 1 - 10 µM (the range of concentrations used for the ASP dye spectra in **Figure S3**, see also our previous work[49]). Hence, assuming an irradiated light beam cross section of 1 cm$^2$, the sample volume probed by the beam corresponds to 1 cm$^3$. With the given solute concentration range of 1 - 10 µM, this means that between $3 \cdot 10^{15}$ to $3 \cdot 10^{16}$ molecules are sampled in a typical ASP measurement of a light-absorbing solute. In contrast, for NSS, we operate in the 100 mM concentration range but with a sample volume of 1.2 fl (or ca. $8 \cdot 10^{-13}$ cm$^3$) only. This translates to $72 \cdot 10^6$ molecules probed in an NSS experiment at 100 mM concentration and corresponds to a staggering 9 orders of magnitude reduction in the number of molecules required. This, in combination with the fact that the NSS sample volume can be further reduced to the attoliter (al) regime by reducing the nanochannel dimensions and/or the channel section used for analysis, highlights the ability of NSS to work with very tiny amounts of sample substance. As a proof for this point, we fabricated a fluidic system identical in structure to the one described above, but now featuring sample nanochannels of 100 nm width and 180 nm depth with a total length of 120 µm. By then splitting the total length of the channel into longer (19 µm) and shorter (3.6 µm) segments and using Brilliant Blue dye across the 5 mM to 100 mM concentration regime, we reduced the sample volume from which the RSID spectra are recorded to 340 al ($20.5 \cdot 10^6$ molecules @ 100 mM and $1.025 \cdot 10^6$ molecules @ 5 mM , **Figure 7a**) and 65 al ($4 \cdot 10^6$ molecules @ 100 mM and $2 \cdot 10^5$ molecules at 5 mM, **Figure 7b**), respectively. Clearly, the spectral fingerprint of the dye is still resolved, also for these tiny sample volumes. A comparison with the RSID spectra taken for the same dye concentrations in the larger nanochannels (**cf. Figure 6b**) reveals that the positions of the spectral features are identical, even though noise level increases and absolute RSID is decreased due to the smaller scattering cross section of the smaller channel and the smaller number of molecules inside said smaller channel (**Figure S6**). Taken all together, these results corroborate the ability of NSS to resolve the distinct spectral fingerprints of dye molecules in the visible spectral for sample volumes in the attoliter to femtoliter range, for concentrations in the 5 mM to 100 mM concentration regime. This corresponds to as few as $2 \cdot 10^5$ probed molecules only, which is 10 orders of magnitude fewer molecules as usually probed in established ASP.

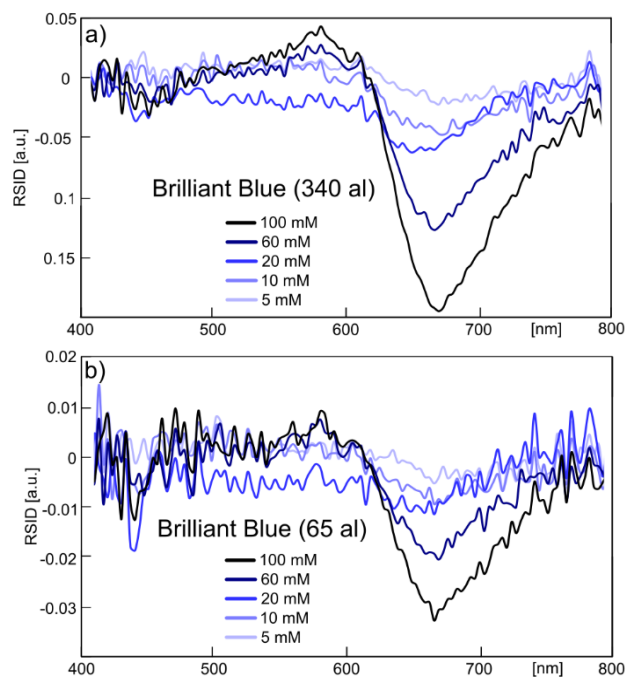

*Figure 7. NSS from attoliter sample volumes.* *a) RSID spectra for different concentrations of Brilliant Blue in water taken from a 19 µm long section of a 100 nm x 180 nm channel, resulting in a sample volume of 340 al. b) Same as a) but here spectra were taken from an only 3.6 µm long section of the nanochannel, reducing the sample volume to 65 al.*

*Concentration dependence*

To further develop the NSS concept with focus on quantitative information that can be extracted, we focus on the concentration dependence of NSS spectra. As mentioned above, the proportionality between the RSID amplitude and the RI of the liquid in the channel (**Equation 1, Equation S1**) enables the quantitative determination of the concentration of a specific solute in a nanochannel. To further explore this opportunity, we plot the RSID amplitude extracted at the wavelength with the strongest response for each compound vs. the predetermined concentration for the five solutes investigated in this study (**Figure 8a, b**), i.e., $H_2O_2$ (RISD at 612 nm, **Figure 8c**), NaCl (RISD at 580 nm, **Figure 8d**), Allura red (RISD at 570 nm, **Figure 8e**) and Fluorescein (RISD at 510 nm, **Figure 8f**), Brilliant Blue (RISD at 665 nm, **Figure 8g**), again using the larger channels with the 1.2 fl sample volume.

The first key observation is that all data points across the 5 different solutes follow a linear relation to a good first approximation since the range of RI change at hand is sufficiently small. This is in good agreement with the prediction in **Figure 1e**. Practically, this means that an NSS fluidic chip can be pre-calibrated for concentration measurements of a specific solute and thus enable accurate measurements of the concentration dependence of solutes in sample volumes in the femtoliter regime and below.

The second key observation is that each solute has a specific slope for this linear relation, which corresponds to the change in scattering intensity per change in concentration at the given wavelength. We find the smallest slope for $H_2O_2$ (-0.01 RSID/M) and the largest one for Brilliant Blue (-7.59 RSID/M), as shown in **Figure 8a-b**. The underlying reason for the different slopes exhibited by different compounds is straightforward to understand for NaCl and $H_2O_2$, since they do not absorption light in the visible regime. Hence, it is only the actual change of the real part of the RI that causes the change in RSID. At an absolute scale, the difference in slope between $H_2O_2$ (-0.01 RSID/M) and NaCl (-0.04 RSID/M) can be explained by considering the Lorentz-Lorenz formula that directly connects RI and the polarizability of the solutes[61]. To this end, in their investigation of the RI of salt water, Aly et al.[61] concluded that the RI of a solution is determined by the ratio of solute and solvent molecules but also by the molar polarizability of each species. Hence, based on the different slopes depicted in **Figure 8a**, it is clear that the molecular polarizability of NaCl in water is higher than that of $H_2O_2$. The reason is the dissociation of NaCl into its charged ionic constituents, while $H_2O_2$ remains molecularly intact, which makes the dissolved Na+ and Cl- ions more responsive the to the external electrical field of light.

Turning to the dyes, which all exhibit strong absorption bands in the visible spectral range, the observed larger slopes have their origin in the change in RI caused by the strong absorption in the visible regime. This means that the slopes depicted in **Figure 8b** are directly connected to the molar extinction coefficient, ε, for each solute, as shown in **Table 1**.

| Dye | RSID slope [RSID/M] | ε [$10^4$ $M^{-1}cm^{-1}$] |
|---|---|---|
| Allura Red | -2.47 | 2.78[62] |
| Fluorescein | -3.43 | 7.69[63–65] |
| Brilliant Blue | -7.59 | 13.0[66] |

*Table 1. Comparison of RISD concentration slopes and the molar extinction coefficient, ε, for the Allura red and Brilliant Blue dyes and for Fluorescein.*

This clear connection between change in scattering intensity and molar extinction further illustrates how compounds can be identified and quantified at the nanoscale using NSS, since a solute with a higher molar extinction coefficient will have a higher amplitude in the RSID spectrum at the same concentration.

As second aspect of relevance here is the lowest concentrations of a specific solute that can be resolved. It becomes clear from the RSID spectra of our five solutes taken at the three lowest concentrations, for each respective system in a 200 nm deep and wide nanochannel, that for the two non-absorbing solutes we are approaching the limit of detection in the 250 – 500 mM range (**Figure 8c, d**), whereas for the dyes, the limit is roughly one order lower, i.e., in the low mM range or below (**Figure 8e-g**). If desired, either increasing the intensity of the irradiated light[55] or increasing the nanochannel cross-section, since the scattered intensity is proportional to channel dimensions (see **Equation 1**), will enable measurements at lower concentrations.

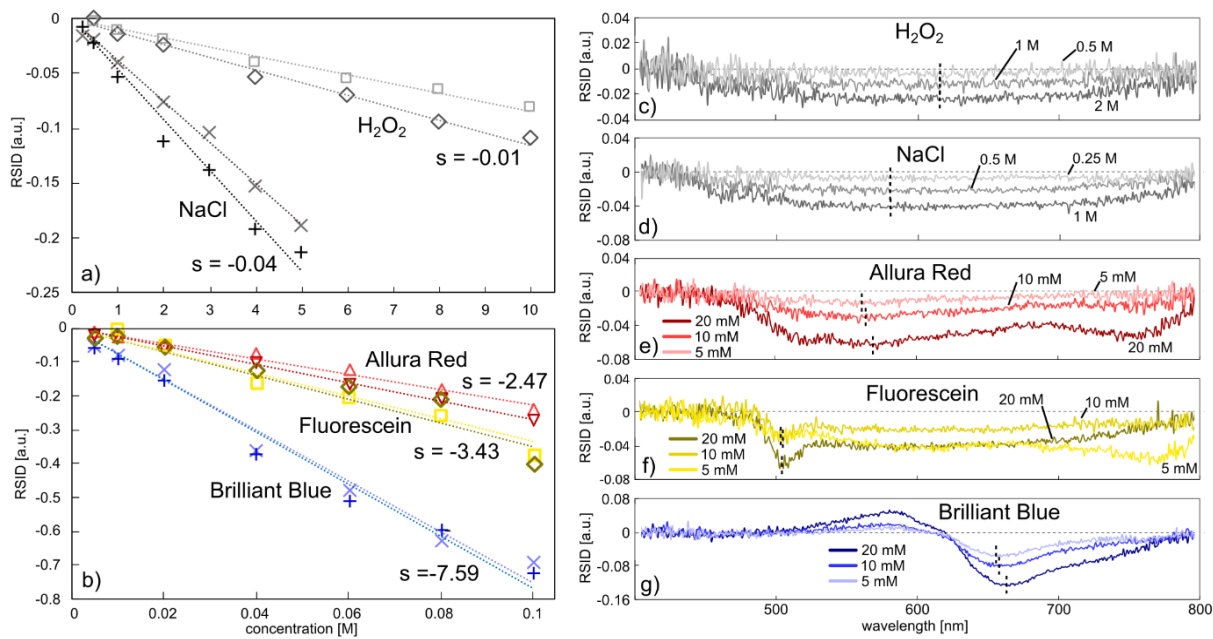

**Figure 8. *Solute concentration dependence.*** *a) Comparison of the colorless solutes $H_2O_2$ and NaCl in terms of RSID change per change in concentration of the solute in the nanochannel for two independent measurement series. The symbols are data points taken at 580 nm (for $H_2O_2$) and 612 nm (for NaCl) where the RSID contrast is largest. The dashed lines are linear fits to the data points whose slope, s, is indicated as average of the two respective measurements. b) Same as a) but for the dyes Allura Red (measured at 570 nm), Brilliant Blue (measured at 665 nm) and Fluorescein (measured at 510 nm). The respective peak positions given where used for all dye concentrations above 20 mM, where an average of 10 data points around this position was taken. For concentrations lower than that, the peak positions seemed slightly shifted, so that the peak position for the evaluation was adapted (vertical dashed lines in c-g). c - g) The RSID spectra for the three lowest concentrations of each solute together with the peak position (black dashed lines) used for extracting the data points in a) and b).*

*Extracting molar extinction coefficients from NSS spectra*

Having established the ability of NSS to deliver quantitative information about solute concentration in the molar to millimolar concentration range in femtoliter and attoliter sample volumes, as the last aspect, we explore the possibility to determine the molar extinction coefficients of the solutes. To do this, we first reconnect to the beginning of this work, where we used the Kramers-Kronig relation (**Equation 4**) to analytically link the change in optical absorption of a solution to a corresponding change in its RI, which in turn is the reason for the scattering intensity difference measured by NSS. Using the same principle, it is possible to calculate a molar extinction spectrum from an RSID spectrum (which reflects the underlying RI spectrum) produced by NSS. The detailed analytical derivation of the mathematical formalism is presented in **SI Section II**. In brief, the reverse calculation of molar extinction coefficients from RSID spectra consist of the following steps: i) Recalling that RSID is the ratio of the sample and reference channel scattering intensities, which means that it also is the ratio of the scattering cross sections of the sample and reference nanochannels (see **Equation S1**). ii) Using **Equation 1** to relate the scattering cross section of the nanochannels to the RIs of water (the solvent here), the solution itself, and the material the nanochannel is embedded in, i.e., $SiO_2$). iii) Calculating the RI spectrum of the solution and fitting a Cauchy-type curve to the obtained RI spectrum to enable subsequent subtraction of this "normal" dispersion from the "anomalous" dispersion, i.e., here the absorption bands of the dye that we are interested in since they eventually translate into the extinction coefficient. To this end, the Cauchy formula is a basic analytical description of the dispersion in a transparent medium[67], whose simplicity facilitates the separation of "normal" and "anomalous" dispersion in a reasonably straightforward way. We note that for cases where the investigated spectral range extends further into NIR-regime, a fit of the Sellmeier-equation[68] may be more appropriate. iv) Using the isolated change in the RI spectrum that stems from anomalous dispersion by the dye absorption bands and applying the Kramers-Kronig formula to calculate an extinction coefficient spectrum. v) Calculating the molar extinction coefficient spectrum based on the known concentration of the solute in the solution. This calculation is relatively straightforward, since all geometry-dependent factors can be cancelled out because reference and sample nanochannels have identical dimensions. However, we note that the quality of the Cauchy-fit to the RI spectrum is important when the molar extinction coefficient spectrum is to be calculated from scattering, since the presence of the dye molecules alone (irrespective of their specific absorption bands) changes the RI of the solution, as shown for the non-absorbing solutes NaCl and $H_2O_2$. The fit therefore needs to include all contributions to the RI that do not stem from anomalous dispersion, such that they can be subtracted. However, this subtraction is challenging in practice since the exact contribution of normal and anomalous dispersion is not known. Hence, the applied Cauchy-fit is to be regarded as a reasonable first approximation to the problem at hand, whose validity is corroborated by the good agreement we obtain in terms of extracted extinction coefficients for all three dyes (**Figure 9** and **Figure S7**). We also note that accurately knowing the RI of the surrounding material (a Borofloat 33 cover glass bonded to the thermal oxide of the silicon wafer in our case) is equally important since it directly determines the amplitude of the RSID spectra. This becomes evident from **Equation 1**, which states that the ratio of the RI of the channel content and the RI of the channel embedding matrix contribute almost quadratically, making NSS as sensitive to the RI *outside* of the channel as to the RI *inside* it.

With these minor reservations, we now apply this formalism to the three dyes, to calculate their molar extinction coefficient spectra from the RSID spectra and compare them with the ones obtained from traditional ASP (**Figure 9a-c**). Overall, we find very good agreement for all three dyes. For a more detailed inspection, we first consider the spectra for Brilliant Blue and see that both the main absorption peak at 630 nm and the shoulder at 590 nm match well in the NSS and ASP spectra (**Figure 9a**). In particular, the agreement in peak amplitude at 630 nm, and thus in absolute molar extinction coefficient for the main dye absorption band, is excellent. Furthermore, we find that the 590 nm shoulder is more pronounced in NSS when compared to the ASP spectrum, which we attribute to the specifics of the illumination spectrum in the NSS experiments that exhibits a peak at this position (cf. **Figure 6a**). The small peaks below 450 nm in **Figure 9a** appear offset by about 20 nm. As the reason, we argue that the peak at 450 nm in the NSS spectrum first and foremost stems from the peak in the irradiated light spectrum (**cf. Figure 6a**) since it overlaps with actual absorption peak of the dye at 412 nm. This convolution of two contributions renders the peak in the NSS extinction coefficient spectrum slightly skewed to shorter wavelengths, which is the reason for the observed apparent offset.

Turning to Allura Red, we find the main absorption peak with both methods at 490 nm, and with excellent agreement in terms of maximum amplitude and thus molar extinction coefficient for the main absorption band (**Figure 9b**). The second peak at 540 nm, distinctly resolved by ASP, is less pronounced in the NSS spectrum. We attribute this primarily to the decreased scattering intensity in this area (cf. **Figure 6a**), but it could also be that the Cauchy-fit is distorting the resulting spectrum slightly, since the exact contribution of the anomalous dispersion to the RI spectrum is unknown, as discussed above. For Fluorescein (**Figure 9c**), the position of the main peak at 500 nm, as well as its amplitude, are in almost perfect agreement for NSS and ASP. The amplitude of the shoulder on the short-wavelength-side of the main peak is slightly out of proportion, again due to the with the low light transmission through the microscope in this spectral range and the corresponding reduction of the S/N.

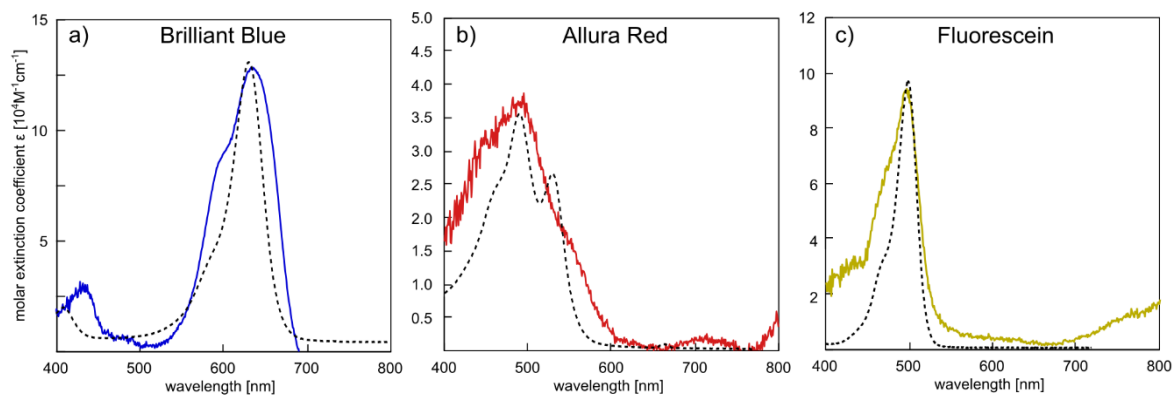

*Figure 9. **Molar extinction coefficient spectra obtained by ASP and NSS.** Molecular extinction coefficient spectra for a) Brilliant Blue, b) Allura Red and c) Fluorescein, as measured using traditional ASP (3.8 µM, 14 µM and 19 µM concentration, respectively - dashed lines) and obtained by NSS (solid lines) via the back-calculation from RISD spectra detailed in **SI Section SII**.*

As the last aspect, we briefly mention possible improvements to NSS as a method for the measurement of the molar extinction coefficient of solutes: i) More exact measurements of the RI of the matrix material that surrounds the nanochannel; ii) Optimization of the fit to the contribution of the normal dispersion to the RI spectrum by, e.g. using a polynomial instead of a Cauchy curve; iii) Reduction of the general noise level in the data by increasing the exposure time, the number of averaged spectra and the intensity of the incident light, or by using a camera with lower background noise for recording; iv) optimizing the microscope setup and spectrometer grating to enable higher light transmission at shorter wavelengths to increase S/N in this range, and thereby better resolve spectral features in this regime.

**Conclusions**

We have introduced Nanofluidic Scattering Spectroscopy, NSS, which measures visible light scattered from a single nanofluidic channel in a spectrally resolved way, as a tool for the spectroscopic investigation of colored and transparent solutions inside nanofluidic nanochannels in continuous flow fashion, and for sample volumes as small as a few tens of attoliters in the millimolar concentration regime, which corresponds to as few as $10^5$ probed molecules. As a further key step beyond the state of the art, we have implemented two independent fluidic systems on the used chip. The first one served as the sample system through which different sample solutions were introduced, whereas the second one constituted an optical reference system filled with water at all times, which we used to compensate for, e.g., fluctuations in irradiated light intensity during a measurement. In this way, we were able to implement a similar concept on a nanofluidic chip as widely used in dual-beam spectrophotometers to establish long-term stable measurement conditions. On the examples of the non-absorbing solutes NaCl and $H_2O_2$, and the dyes Brilliant Blue, Allura Red and Fluorescein, we subsequently demonstrated that their spectral fingerprints can be obtained with good accuracy and that solute concentrations inside a single nanochannel can be determined based on NSS-spectra from femto- to attoliter sample volumes. Furthermore, by applying a reverse Kramers-Kronig transformation to measured NSS-spectra of solutes inside a single nanochannel, we demonstrated that the molar extinction coefficient spectrum of the solute can be extracted with very good agreement with literature values, thereby enabling the identification of solutes based on a fundamental material constant.

Looking forward, our findings advertise NSS as a versatile tool for the spectroscopic analysis of solutes in situations where nanoscopic sample volumes, as well as continuous flow measurements, are of importance. An example for such a situation is single particle catalysis inside nanofluidic channels, which has the aim to resolve catalytic conversion on the surface of a nanoparticle localized inside the nanochannel by preventing excessive dilution of the reactant products due to the tiny volume of said nanochannel[45,49,69,70] To this end, for this application, we have already demonstrated the key importance of limiting the detection to a fraction of a nanochannel[46] and argue that adding the spectroscopic dimension enabled by NSS further leverages the potential of single particle catalysis in nanofluidic systems as a whole. Furthermore, since it is possible to realize various concentration, temperature and pressure scenarios inside nanofluidic systems, we envision that the NSS concept also may find application in homogeneous catalysis, where, e.g., the absorption of photosensitizers[71] could then be investigated on only fractions of the usual sample volume, or in biological sample sequencing and medical drug development where costly molecules initially are available in tiny amounts and volumes only. To reduce the total sample volume required in such experiments, i.e., the volume equivalent to the entire fluidic system connecting to the nanochannels and not only the nanochannels themselves, the micro-and nanofluidic concepts employed offer ample opportunities as they easily can be downscaled in their entirety.

**Methods**

*Instruments*

The exchange and flow of solutions in/through the nanofluidic chip was realized by applying pressure to the inlet reservoirs of the chip holder with a Fluigent MFCS-EX pressure controller. The dark-field microscopy images were taken on a Nikon Eclipse LV150N upright microscope equipped with a Nikon 50x ELWD dark-field objective. Illumination was provided by a Thorlabs Solis-3C LED light source. The light scattered from the nanochannels was directed into an Andor spectrometer (SR-193I-A-SL) that had a 150 l/mm grating installed. The spectrally resolved light was recorded with a Andor Newton (DU920P-BEX2-DD) camera attached to the spectrometer, which binned the four areas (upper and lower reference channels and upper and lower signal channels) to a total of four spectra, which were used for subsequent data evaluation. The spectra were recorded with 2 s exposure time and consisted of 10 accumulated pictures. For Figure 5 and 6, additional spectra of the incident light were recorded using an Avantes AvaSpec-1024 fiber spectrometer. ASP spectra of the dyes were recorded on a Varian Cary 50 Bio UV-Vis spectrophotometer. The SEM images of the channel and its cross-section were recorded on a Zeiss Supra 55VP scanning electron microscope.

*Sample preparation*

All dyes (Allura Red, Brilliant Blue, Fluorescein) were bought from Merck as solid material and diluted into the required concentrations by mixing the dyes with ultrapure water (Milli-Q IQ 7000 water purification, Merck). A similar procedure was carried out for NaCl and $H_2O_2$, where the latter one was acquired as solution ($H_2O_2$, 35% w/w in $H_2O$, Merck). Injection into the fluidic chip holder was done with syringes and blunt needles (Braun).

*Fluidic chip fabrication*

The micro and nanofluidic chips used in the experiments were fabricated in the clean room facilities of MC2 at Chalmers in Gothenburg. Each chip consisted of a single 4-inch silicon wafer with a thermal oxide layer into which the fluidic structures were etched. The general procedure is described in detail in our earlier work by Levin et al[49]. As a short summary: The 4-inch (100) silicon wafers were cleaned with Standard Clean 1, followed by a 2% HF bath and Standard Clean 2. Growing of the thermal oxide layer was carried out in wet atmosphere at 1050 °C until a thickness of 250 nm was reached. Etching of the nanochannels was done via fluorine-based reactive ion etching (RIE) after they have been patterned in a resist layer by electron beam lithography. Subsequently, the microchannels were etched into the thermal oxide by the same method, the photoresist was patterned here with direct laser lithography. As a final step, the substrates were cleaned with Standard Clean 1, together with a 175 µm thick Borofloat 33 glass wafer to be used as lid of the fluidic system. This glass lid has been equipped with inlet holes (sandblasting) to match the liquid reservoir on the nanofabricated wafer. The surfaces of both wafers were then treated with $O_2$ plasma (50 W, 250 mTorr) to enable pre-bonding of the glass cover lid to the wafer with the fluidics. The subsequent fusion bonding was carried out at 500 °C for 5h. After bonding the 4-inch fluidic wafer was cut into shape to fit the chip holder.

**Supporting Information**

This material is available free of charge via the Internet at http://pubs.acs.org.

Schematic of the experimental setup, Nanochannel scattering cross section dependence on wavelength and RI, Spectra of Brilliant Blue, Allura Red and Fluorescein during the analytical transformation from absorption spectra to scattering spectra, Intrinsic scattering spectra and evaluation strategies, Spectra of the light used during the experiment, Comparison of spectra from different channel sizes, Calculation of the molar extinction coefficient from measured RSID spectra. (PDF)

The underlying data for this publication is available at Zenondo, 10.5281/zenodo.10848089.

**Corresponding Author**

Christoph Langhammer − Department of Physics, Chalmers University of Technology, SE-412 96 Gothenburg, Sweden; orcid.org/0000-0003-2180-1379; Email: clangham@chalmers.se

**Authors**

Björn Altenburger − Department of Physics, Chalmers University of Technology, SE-412 96 Gothenburg, Sweden; orcid.org/0009-0003-0600-4635

Joachim Fritzsche − Department of Physics, Chalmers University of Technology, SE-412 96 Gothenburg, Sweden; orcid.org/0000-0001-8660-2624

**Author Contributions**

The manuscript was written through contributions of all authors. All authors have given approval to the final version of the manuscript.

**Acknowledgements**

This research has received funding from the Swedish Research Council (VR) Consolidator Grant project 2018-00329 and the European Research Council (ERC) under the European Union's Horizon Europe research and innovation program (101043480/NACAREI). Part of this work was carried out at the Chalmers MC2 cleanroom facility and at the Chalmers Materials Analysis Laboratory (CMAL). We also acknowledge fruitful discussions with Dr. Barbora Špačková.

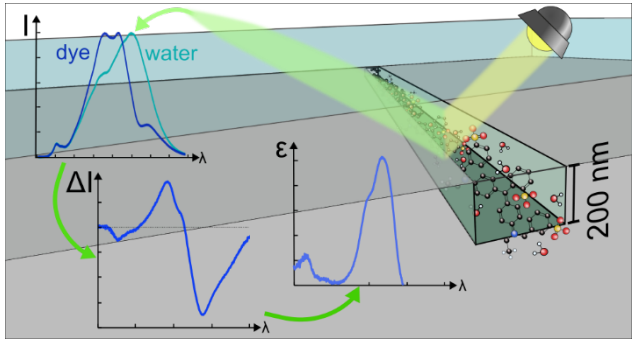

For Table of Contents Only

# Supplementary Material for

# Visible Light Spectroscopy of Liquid Solutes from Femto- to Attoliter Volumes inside a Single Nanofluidic Channel

*Björn Altenburger[1], Joachim Fritzsche[1] and Christoph Langhammer[1]\**

[1]Department of Physics, Chalmers University of Technology; SE-412 96 Gothenburg, Sweden

*Corresponding author: clangham@chalmers.se

**Section I: Supplementary figures**

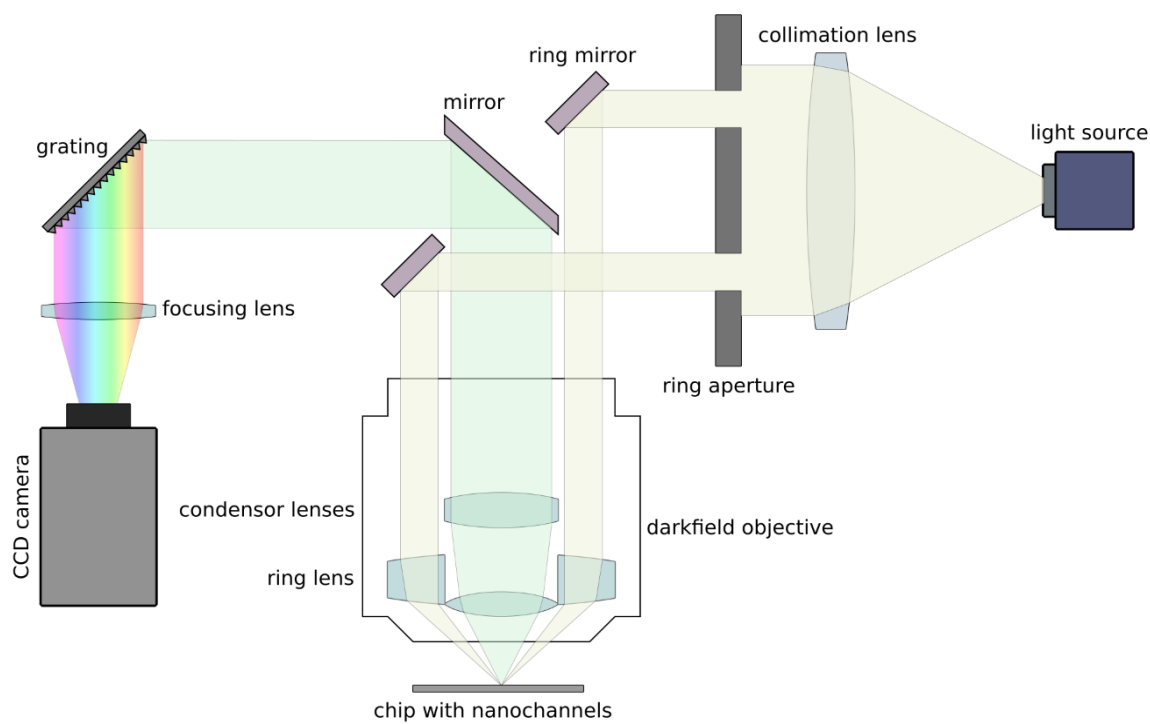

**Figure S1.** *Schematic of the experimental setup.* The illuminating polychromatic white light is emitted by a broad-spectrum LED-lamp (Thorlabs Solis 3C) and collimated into the microscope (Nikon Eclipse LV150N). Here, it is partly blocked by a ring aperture that transmits only a ring-shaped section of the light onto a ring-shaped mirror, from where it is directed into the dark-field objective (TU Plan ELWD 50x/0.6 B OFN25 WD 11). As the objective is specifically built for dark-field illumination, it has a separate light path for the incident light, at the end of which a ring-shaped lens focuses the light in the focus spot of the imaging path. Here, a set of lenses collects the scattered light into a collimated beam that then is directed into the spectrometer (Andor SR-193I-A-SL), where a grating (150 l/mm) separates the individual wavelength components. The spectra are finally recorded by a CCD camera (Andor Newton DU920P-BEX2-DD).

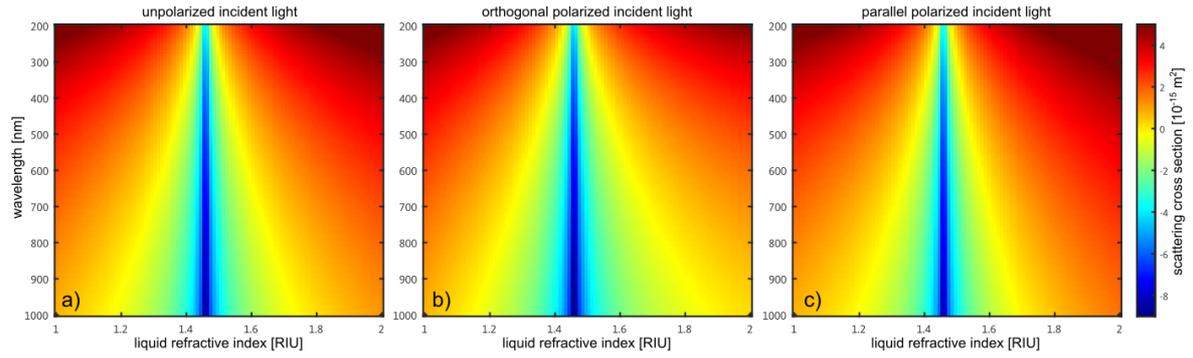

**Figure S2.** *Nanochannel scattering cross section dependence on wavelength and RI.* The three panels show how the scattering cross section, of a 200 nm by 200 nm nanochannel embedded in $SiO_2$ (for which the RI is calculated for each wavelength according to Malitson et. al.[1]) depends on the wavelength of the incident light and the RI of the medium that fills the channel, for different polarizations of the incident light respectively. The color code axis is given in logarithmic units with base ten. a) For unpolarized light, the scattering cross section of the channel ($\sigma_{channel}$) shows a minimum around the RI of $SiO_2$, i.e. $n = 1.459$ at 600 nm, since in this case the channel and its surrounding medium form an optically uniform body. Deviations from this value lead to an increase in the $\sigma_{channel}$ for all wavelengths. For shorter wavelengths, $\sigma_{channel}$ increases more rapidly to higher values. For RIs of the channel that are larger than the surrounding material, $\sigma_{channel}$ increases slightly more than when compared to the same RI difference for a channel with lower RI (see also **Figure 1e** in the main text). b) For light that is polarized orthogonal to the channel, the overall result is the same as in a), with the exception that $\sigma_{channel}$ is not as large as in the unpolarized case. c) For light that is polarized parallel to the nanochannel the overall $\sigma_{channel}$ is larger than in the other two cases. Especially remarkable here is the increase of $\sigma_{channel}$ when the channel has a higher RI than the surrounding material, which becomes even more significant when considering shorter wavelengths.

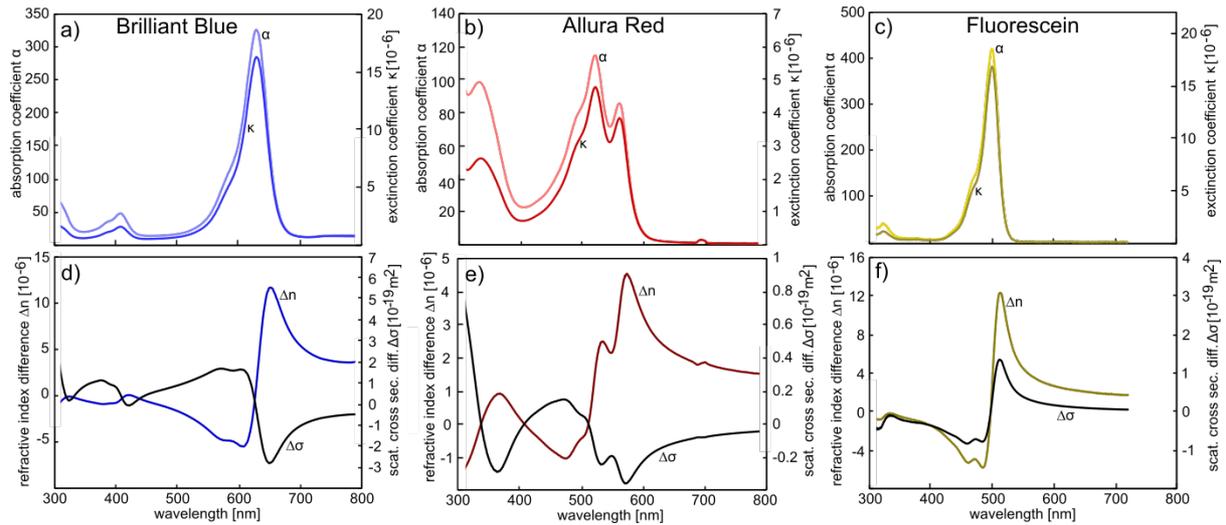

**Figure S3.** *Spectra of Brilliant Blue, Allura Red and Fluorescein during the analytical transformation from absorption spectra to scattering spectra.* a) to c) The absorbance spectra as measured with absorption spectrophotometry (ASP) using an Varian Cary 50 Bio instrument are here shown as absorption coefficient spectra, which can be calculated by dividing the measured absorbance values by the optical path through the sample solution (here 1 cm) and by log(e). The subsequent calculation of the extinction coefficient κ is done with **Equation 3** in the main text. d) to f) Using κ in the Kramers-Kronig relation (**Equation 4** in the main text) yields the change of the real part of the refractive index (RI) caused by the respective absorption features of each type of dye molecule in the solution. Applying then **Equation 1** from the main text delivers the scattering cross section, $\sigma_{channel,solution}$, of a nanochannel filled with the corresponding dye solution and corresponding concentration. Subtracting from $\sigma_{channel,solution}$ the scattering cross section, $\sigma_{channel,water}$, of a water-filled channel yields the difference in scattering cross section, Δσ, which is proportional to the observed difference in scattering intensity (RSID). The calculated spectra for Δσ are used as a qualitative benchmark in the comparison with the experimental RSID spectra since the dye concentrations used in the nanochannels are too high to be measured with ASP.

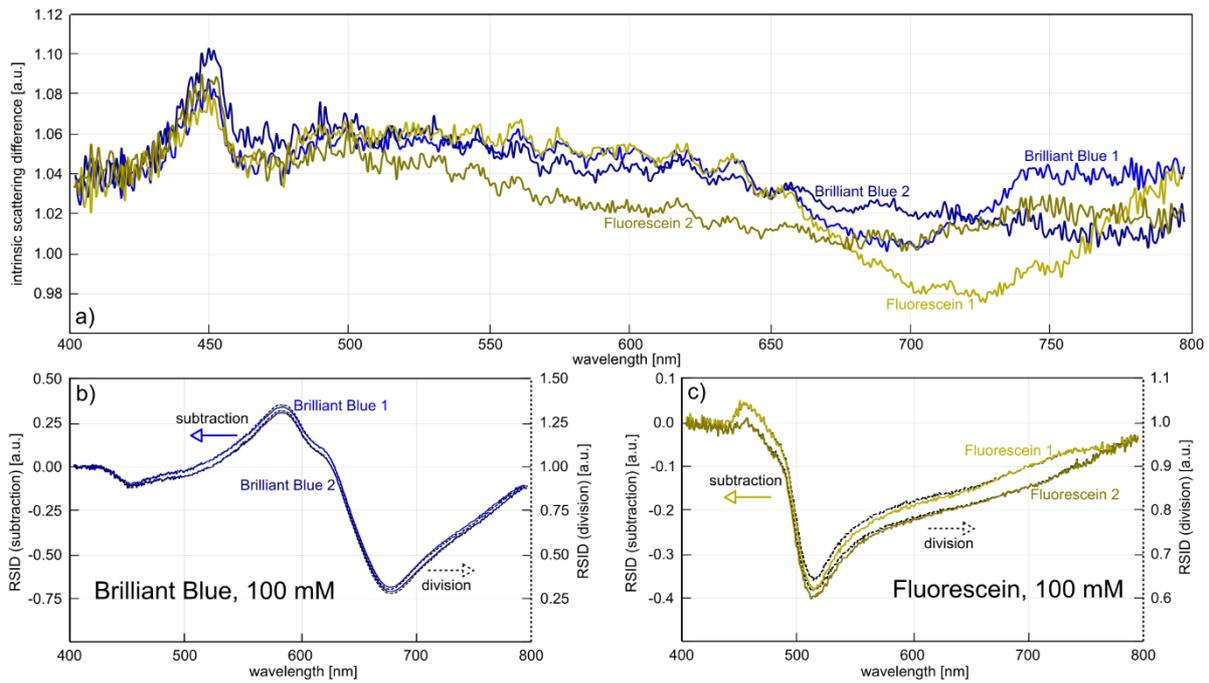

**Figure S4.** *Intrinsic scattering difference and comparison of evaluation strategies.* *a) Intrinsic scattering difference spectra for two independent measurement series of the dyes Brilliant Blue and Fluorescein (100 mM) as used for **Figure 6b,d** and **Figure 8b**. The intrinsic scattering difference is here defined as the ratio of the scattering intensity of a water-filled sample channel and the scattering intensity of its corresponding water-filled reference channel. The colors indicate to which of the two subsequent RSID measurements of Brilliant Blue (blue) and Fluorescein (yellow) they belong, which are shown in b) and c) respectively. b) RSID spectra for two independent measurements of a 100 mM Brilliant Blue solution in a 200 nm by 200 nm channel. The solid lines show the result when the intrinsic scattering difference is subtracted (as described in the main text, **Figure 4**), while the dashed lines are the result when instead division by the intrinsic scattering difference is used. c) Same as for b), however, for a 100 mM Fluorescein solution.*

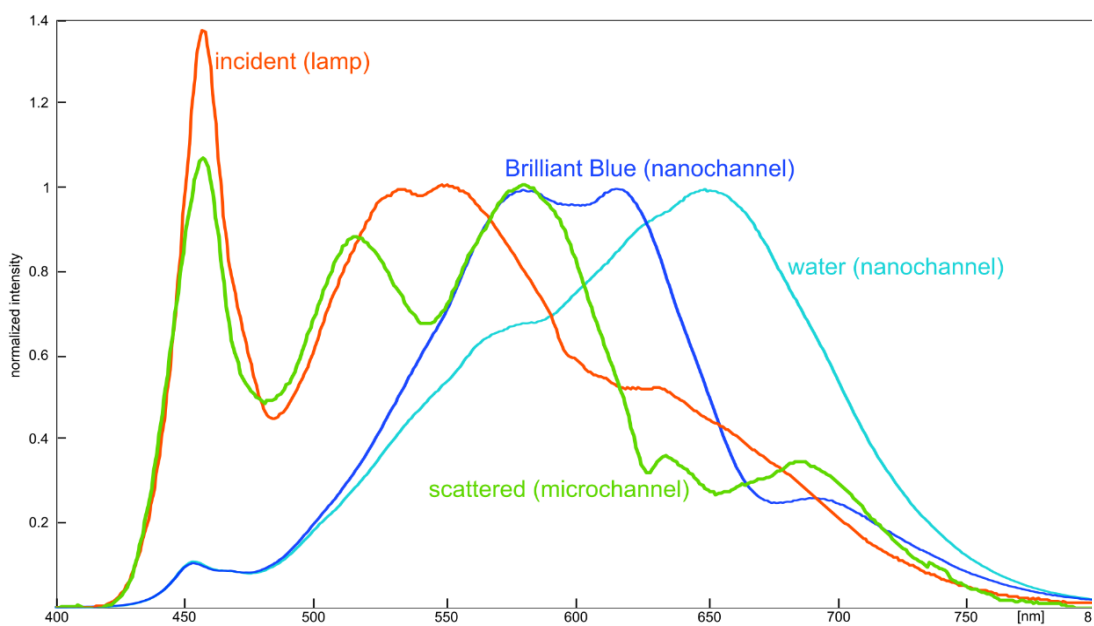

**Figure S5.** *Spectra of the light used during the experiment, scaled to their maximum between 500 nm and 700 nm.* The initial incident light spectrum (red line) measured with an Avantes AvaSpec-1024 spectrometer is emitted from a Thorlabs Solis-3C LED light source. It undergoes several changes regarding the shape of the spectrum during its course through the experimental setup. The light source is a broad-band LED-lamp, that has its main intensity centered around 550 nm but also an additional strong peak at 450 nm. After being focused on the fluidic system, the light scatters on the fluidic channels (green line, scattering from a microchannel wall for intensity reasons, recorded with an Avantes AvaSpec-1024 spectrometer). Here we see that the spectral shape of the lamp is not maintained. Most remarkable is the dip that has appeared at 540 nm but also the peak at 450 nm has decreased in relative size. We assume that thin-film interference in the $SiO_2$ layer of the fluidic chip causes these changes. After being scattered from a water-filled nanochannel (cyan line, measured with a Andor SR-193I-A-SL spectrometer and a Andor Newton DU920p-BEX2-DD camera), the shape of the spectrum has again changed, now being more bell-shaped with a maximum at 650 nm. We see the main cause for this in the sensitivity of the grating spectrometer and the camera. The peak at 450 nm is now greatly diminished when compared to the original spectrum. When the nanochannels are now filled with a dye (Brilliant Blue as example), the spectrum changes again in shape according to the change in absorption and RI as explained in the main text. All spectra have been normalized to their maximum value in the wavelength range over 500 nm.

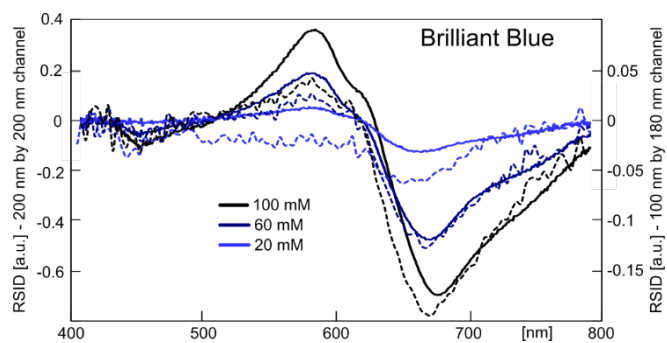

**Figure S6.** *Comparison of RSID spectra measured from nanochannels with different dimensions.* RSID spectra for three different Brilliant Blue concentrations measured in a nanochannel with 200 nm x 200 nm cross section (continuous lines) a nanochannel with 100 nm x 180 nm cross section (dashed lines).

## Section II: Supplementary derivations

Calculation of the molar extinction coefficient from measured RSID spectra

In the main text, we explained how the expected relative scattering intensity difference (RSID) spectrum can be calculated from the absorbance spectra of the respective dyes. Here, we will outline the reverse process leading to **Figure 9** in the main text. As a first step, we need to establish the connection between RSID and scattering cross section of the nanochannel, $\sigma_{channel}$. The scattered power, $P_{scat}$, is the product of incident intensity, $I_{incident}$, and $\sigma_{channel}$. The RSID as shown in the main text is the scattering spectra recorded from the solution-filled channels divided by the scattering spectra from the water-filled channel but then also with the *intrinsic difference spectrum* subtracted. To amend this, we add an ideal *intrinsic difference spectrum* for a ratio, which is 1, to the RSID spectra (see **Figure S7a-c**). Here, the shaded area below 425 nm for all graphs indicates the wavelength range where the intensity of the incident light and the sensitivity of the spectrometer and camera is insufficient to conduct a measurement of RSID.

$$RSID + 1 = \frac{P_{scat,solution}}{P_{scat,water}} = \frac{I_{incident}\sigma_{channel,solution}}{I_{incident}\sigma_{channel,water}} = \frac{\sigma_{channel,solution}}{\sigma_{channel,water}} \quad \text{Equation S1}$$

As second step, we consider the scattering cross section, $\sigma_{channel}$, of a nanochannel as given by **Equation 1** in the main text, but as a simplification consider here only incident light that is polarized parallel to the nanochannel, as it is the main contributor to the scattering intensity (see **Figure 1e**).

$$\sigma_p = \frac{A_\emptyset^2 k^3 L}{4}(m^2 - 1)^2. \quad \text{Equation S2}$$

With this expression at hand, we can write the RSID as a ratio of scattering cross sections, were the geometry and wavelength dependent pre-factor vanishes, and were we substitute $m$ again as the ratio of RIs of the solution in the channel and the SiO$_2$ the channel is embedded in.

$$RSID + 1 = \frac{\sigma_{channel,p,solution}}{\sigma_{channel,p,water}} = \frac{\left(\left(\frac{n_{solution}}{n_{SiO2}}\right)^2 - 1\right)^2}{\left(\left(\frac{n_{water}}{n_{SiO2}}\right)^2 - 1\right)^2} \quad \text{Equation S3}$$

Solving this equation for $n_{solution}$ provides four solutions, of which we will continue with the following as it is not resulting in negative values for $n_{solution}$ and applies to the case where $n_{water} < n_{SiO2}$.

$$n_{solution} = \sqrt{\sqrt{(RSID + 1)(n_{water}^2 - n_{SiO2}^2)^2} + n_{SiO2}^2} \quad \text{Equation S4}$$

Using here the literature values for the RIs for water[2] and SiO$_2$[1], we arrive at the RI spectra for the respective solutions, as shown in **Figure S7d-f**. It is seen clearly that the dye solutions have RIs larger than water ($n_{H2O} = 1.333$ at 600 nm). The absorption bands of the dyes appear as deviations from the smooth, Cauchy-type curve (yellow dashed line) fitted to the RI spectra. In the presented calculation, it is assumed that the respective dye solutions do not have any absorption in this regime and follow a normal Cauchy dispersion. This assumption enables the fitting of a Cauchy-type function in **Figure S7d-f**. This fit is necessary for the next step of our transformation of RSID to molar extinction coefficient, since only the RI features caused by the absorption bands of the dyes are of interest in the Kramers-Kronig relation (and the corresponding inverse relation) and since the RI spectrum, in addition to the features related to the absorption bands, contains contributions that are not related to absorption, which we need to subtract that part using the Cauchy-fit to estimate it.

The Cauchy-formula is commonly used as a simple description of the normal dispersion relation of transparent media in the visible regime, making it useful to us here to separate normal from anomalous dispersion (which is the extinction of the dye). To do so, we need to note that due to the fact that the Kramers-Kronig relation contains an integral, the whole RI spectrum experiences a change because of the anomalous dispersion, which is most pronounced for the longer wavelengths[3]. The Cauchy-fit is therefore based on the section of the RI curve that is below 400 nm. Furthermore, to include the overall increase of the RI for all wavelengths as mentioned above, we estimated this contribution to be 0.01 RI units and subtracted it from the initial fit (grey dashed line in **Figure S7d-f**) such that the extracted contribution of the normal dispersion (grey line) happens to be lower than the baseline of the calculated RI spectrum of the dyes. This estimate seems to be reasonable, as it is in line with our understanding of the Kramers-Kronig relation for the normal and anomalous dispersion of a dye solution and reproduces the expected result sufficiently well.

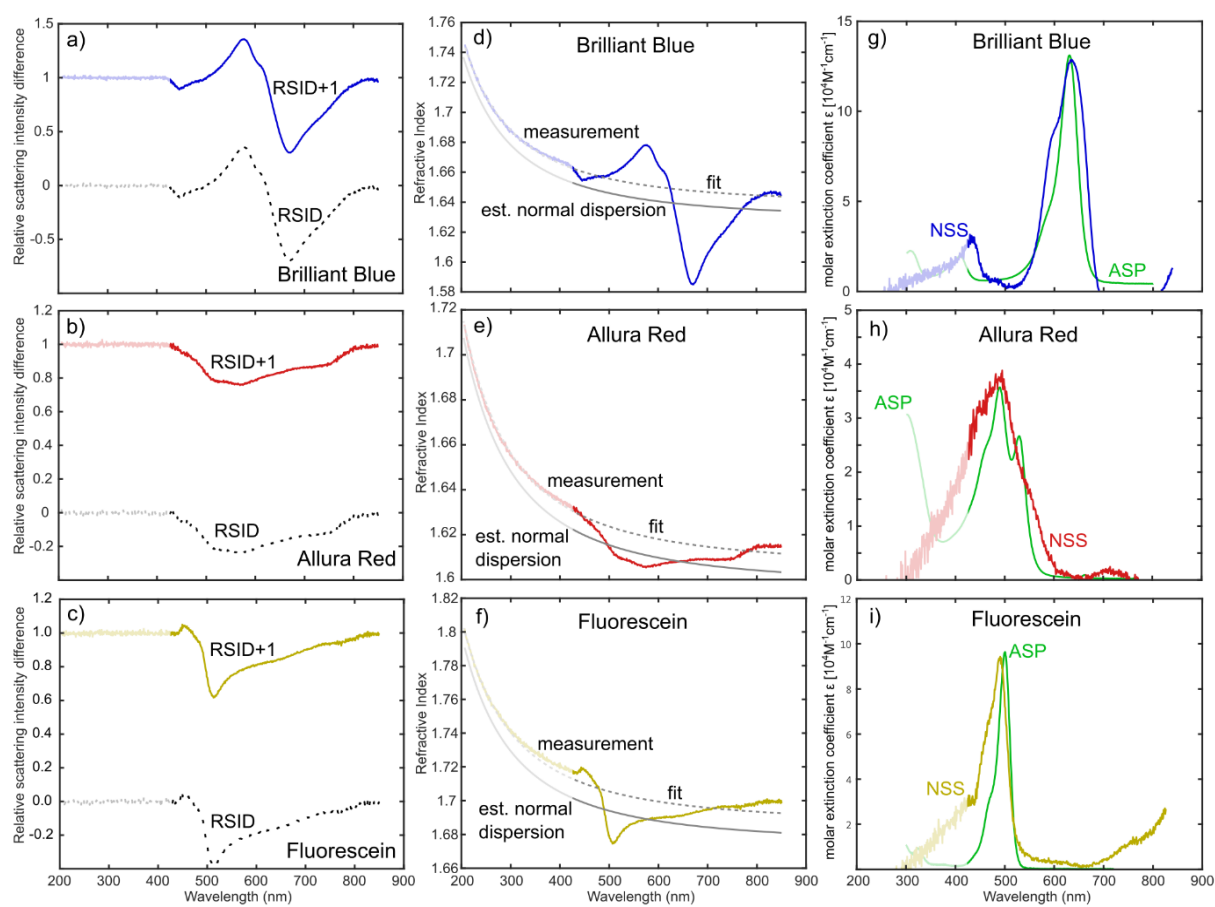

**Figure S7.** *Translation of RSID to molar extinction coefficient. The shaded sections of the spectra indicate the wavelength regime where low incident light intensity and insufficient spectrometer and camera sensitivity prevent a measurement of RSID. a)-c) The RSID spectra from the experiment (**cf. Figure 6b-d**, dashed lines) for a 100 mM Brilliant Blue, Allura Red and Fluorescein solution, respectively. The colored solid lines are RSID+1 spectra and represent the ratio of scattering cross sections as laid out in **Equation S1**. d)-f) Using **Equation S4**, the RSID (which is the ratio of scattering cross sections) can be translated into the RI spectrum of the respective dye solution (colored lines). To extract the part of the RI spectrum that describes normal dispersion and that is not associated with the absorption bands of the solute, a Cauchy-type fit (grey dashed line) is applied to the RI spectrum and the estimated contribution of the normal dispersion (grey line) is later subtracted from the RI curve of the solution. g)-i) Molar extinction coefficient spectra for the three dyes obtained by reverse Kramers-Kronig transformation (colored lines, **Equation S5**) plotted together with the molar extinction coefficient spectra obtained for the same dyes using traditional ASP (green lines). We note the generally good agreement between NSS-based and ASP spectra. The unphysical negative molar extinction coefficients obtained by converting the Brilliant Blue NSS data (g) between 700 nm and 800 nm, comes*

*most probably from the long-wavelength contribution of the anomalous dispersion to the normal dispersion that is not represented in the Cauchy-fit The molar extinction coefficients also become negative for all dyes below 300 nm, but here we assume that the high noise and low intensity of the scattered signal is responsible for this unphysical deviation.*

The inverse Kramers-Kronig relation for the real and imaginary part of the RI (of which the imaginary part correspond to extinction) is nearly identical in structure to **Equation 4** in the main text, except for a minus sign in front and can be used to calculate the wavelength-dependent extinction coefficient as

$$\kappa(\lambda) = -\frac{2}{\pi} \mathcal{P} \int_0^{-\infty} \frac{\Delta n(\lambda')}{\lambda' \left(1 - \left(\frac{\lambda'}{\lambda}\right)^2\right)} d\lambda' \qquad \text{Equation S5}$$

As the very last step, we can then convert the obtained extinction coefficient spectrum, $\kappa(\lambda)$, into the molar extinction coefficient spectrum, $\varepsilon(\lambda)$ by using the known concentration of $c = 100\ mM$ in **Equation S6** for each solution.

$$\varepsilon(\lambda) = \frac{4\pi\kappa(\lambda)}{\lambda\ c} \qquad \text{Equation S6}$$

The final result of this calculation is shown in **Figure S7g-i**, together with the molar extinction coefficients for the respective dyes that have been measured using ASP. The agreement is relatively good regarding the main peak positions and the maximum value of the molar extinction coefficient $\varepsilon$ for all three dyes. The values for $\varepsilon$ do deviate from the ASP spectra slightly but we identify the main reason for this in the separation of the normal and anomalous dispersion as explained above. The Cauchy-fit of the RI spectrum for each respective dye solution may not represent the actual underlying normal dispersion (associated with the real part of the RI of the solution), as it may not fully represent all involved influences of the dye extinction on the total RI. For the shorter wavelengths below 300 nm, poor signal quality due to low scattering intensity can also be considered as reason for these deviations. Furthermore, the RI of the surrounding $SiO_2$ plays a critical role during the calculation, since even a small change in the chosen value has a sizable impact on the final result, since NSS is as sensitive to the RI in the channel as to the RI outside of the channel, as laid out by **Equation 1** in the main text.